\documentclass[prl,superscriptaddress,twocolumn,amsmath,amssymb,floatfix]{revtex4-1}
\usepackage{graphicx}
\usepackage[FIGTOPCAP]{subfigure}
\usepackage[usenames]{color}
\usepackage{amssymb}
\usepackage{hyperref}

\newcommand{\ket}[1]{\lvert#1\rangle}

\begin{document}
\title{Machine learning many-body localization:  Search for the elusive nonergodic metal}
\author{Yi-Ting Hsu}
\email{ythsu@umd.edu}
\affiliation{Condensed Matter Theory Center and Joint Quantum Institute, University of Maryland, College Park, MD 20742, USA}
\author{Xiao Li}
\email{lixiao@umd.edu}
\affiliation{Condensed Matter Theory Center and Joint Quantum Institute, University of Maryland, College Park, MD 20742, USA}
\author{Dong-Ling Deng}
\affiliation{Condensed Matter Theory Center and Joint Quantum Institute, University of Maryland, College Park, MD 20742, USA}
\affiliation{Institute for Interdisciplinary Information Sciences, Tsinghua University, Beijing, China, 100084}
\author{S. Das Sarma}
\affiliation{Condensed Matter Theory Center and Joint Quantum Institute, University of Maryland, College Park, MD 20742, USA}

\date{\today}

\begin{abstract}
The breaking of ergodicity in isolated quantum systems with a single-particle mobility edge 
is an intriguing subject that has not yet been fully understood. 
In particular, whether a nonergodic but \textit{metallic} phase 
exists or not in the presence of a one-dimensional quasi-periodic potential is currently under active debate.
In this Letter, we develop a neural-network based approach to investigate the existence of this nonergodic metallic phase in a prototype model using many-body entanglement spectra as the sole diagnostic. 
We find that such a method identifies with high confidence the existence of a nonergodic metallic phase {in the mid spectrum} at an intermediate {quasiperiodic} potential strength. 
Our neural-network based approach shows how supervised machine learning can be applied not only in locating phase boundaries, but also in providing a way to definitively examine the existence or not of a novel phase.
\end{abstract}

\maketitle

\textit{Introduction---}
Investigating the properties of eigenstates in isolated quantum many-body systems is essential for understanding dynamical phases and their transitions, and  even more importantly, the very question of thermal equilibrium in quantum statistical mechanics. 
In the noninteracting limit, the single-particle orbitals of a fermionic system throughout the energy spectrum can be all localized~\cite{AndersonLocalization}, all extended, or exhibit a single-particle mobility edge (SPME)~\cite{mott1990metal,Soukoulis_Economou,SPME02,SPME03,Griniasty_Fishman,
SPME04,SPME05,SPME06,Boers_2007,SPME07,SPME08,GAASPME} separating localized and extended states.
The SPME is in fact thought to be the generic situation for three-dimensional disordered systems.
Moreover, the existence of an SPME in an incommensurate one-dimensional (1D) system 
has recently been predicted and experimentally observed in a quasiperiodic optical lattice~\cite{SPME_GAA_Theory_2017,SPME_GAA_Experiment_2018}. 

In the presence of interactions, we can further introduce the notion of ergodicity for a many-body eigenstate since a closed quantum system can thermalize according to the eigenstate thermalization hypothesis (ETH)~\cite{ETH1,ETH2,Rigol2008}. 
Since the {critical} energies for the localization and thermalization transitions do not necessarily coincide, more complicated phases can occur other than all eigenstates being many-body localized (MBL)~\cite{MBL01,MBL02,MBL03,MBL04,MBL05,MBL06,Iyer2013Many-body,MBL09,MBL13,MBL14,Imbrie2016, 
MBL08,MBL10,MBL11,MBL12,NEM_PRL_Mukerjee,XiaoPeng_PRL,XiaoPeng_PRB,MBL15,MBL16,NEMdisorderspin,MBL17,MBL18}, i.e., localized and nonergodic, or all obeying ETH, i.e., extended and ergodic. 
In particular, recent numerical studies have found that in systems subject to a family of incommensurate potentials that exhibit SPME, there exists a finite energy window wherein the eigenstates are nonergodic but extended~\cite{XiaoPeng_PRL,XiaoPeng_PRB,NEM_PRL_Mukerjee}. 
Such an intriguing intermediate phase was subsequently named the \emph{nonergodic metal} (NEM)~\cite{XiaoPeng_PRL}. 

The common strategy taken by prior studies to identify NEM~\cite{XiaoPeng_PRL,XiaoPeng_PRB,NEM_PRL_Mukerjee} was to detect localization and ergodicity by different diagnostics. 
This was necessary since different phases are naturally more sensitive to different diagnostics, which is also true in the experimental studies of MBL~\cite{Schreiber_2015,Bordia_2016,Choi_2016,Smith_2016}. 
The problem with this strategy is that these \emph{ad hoc} different diagnostics may not necessarily be equivalent with respect to their sensitivity to various phases. 
For instance, while entanglement entropy and the variance in local particle number fluctuations were used to diagnose localization and ergodicity in Ref.~[\onlinecite{XiaoPeng_PRL}], the inverse participation ratio and the return probability were used in Ref.~[\onlinecite{NEM_PRL_Mukerjee}] along with several other diagnostics. 
Since the energy window for NEM is set by the two transition energies corresponding to the two distinct diagnostics, the phase space or even the existence of NEM itself can largely depend on the combination of the diagnostics used, which is unsatisfactory. 
In order to definitively establish NEM as a phase with a \emph{finite phase space} in the phase diagram, it is imperative to develop an approach that can distinguish MBL, NEM, and the thermal states (ETH states) using a \emph{single diagnostic}.

Entanglement spectrum (ES)~\cite{Haldane2008PRL} is an appealing choice in this context 
because of the following reasons. First, 
the ES contains more information about the eigenstates than entanglement entropy due to the absence of the tracing procedure. 
Second, recent studies have identified ES as a sensitive probe for MBL and ETH phases~\cite{ESlevel_MBLTH,MLES_MBL_Titus,ML_ES_MBLTH,YangPRL2015,powerlawES}. 
However, the complexity of the spectral pattern in ES makes it practically difficult to extract relevant features for discerning eigenstate properties. 

Machine learning, a powerful tool for complex pattern recognition, has recently been introduced to condensed matter physics and raised tremendous interest in the community~\cite{MelkoNphysIsing,van2017Learning,Chng2017Machine,Yoshioka2017Learning,Carleo2016Solving, DLD2017PRB,Wang2016Discovering,
Arsenault2015Machine,Deng2017Quantum,
Wetzel2017Unsupervised,Hu2017Discovering, 
Gao2017Efficient,Huang2017Neural,Chen2018Equivalence,Cai2017Approximating,
Schindler2017Probing,Broecker2017Quantum,Nomura2017Restricted}. 
In particular, supervised learning has been used as a successful numerical tool to study various phases and their transitions~\cite{MelkoNphysIsing,van2017Learning,Chng2017Machine,Yoshioka2017Learning}. 
One such application is to identify the phase boundaries throughout the parameter space using a neural network (NN) classifier trained with data obtained from well-known limits deep in each phase~\cite{MelkoNphysIsing,MLtopoKim,MLES_MBL_Titus,ML_HubbardSign_Trebst}. 
Such an approach, however, relies on \emph{a priori} knowledge of all existing phases in the parameter space, which is not always available. 
To the best of our knowledge, studies in this direction have so far been limited to models where the existence of all phases are well established without controversies, which is not suitable for our goal of investigating the presence of the controversial NEM phase whose very existence as an intermediate phase between ETH and MBL remains open.

In this Letter, we develop a general NN approach targeting a different goal from that of conventional supervised learning---to examine the existence of a controversial phase. Using this novel approach, we investigate the dynamical phases in a prototype incommensurate 1D lattice model ~\cite{XiaoPeng_PRL}, emphasizing the existence of the NEM phase. Using ES as the input data, we show that a three-layer NN is able to unambiguously identify a distinct new phase between MBL and ETH phases with high confidence. 
Our results provide the strongest numerical evidence so far for the existence of a new phase in incommensurate systems that is likely the predicted NEM in Ref. \onlinecite{XiaoPeng_PRL}, 
and also usher in a general machine learning based new technique for identifying novel phases of matter which may not be accessible by conventional techniques. 

\begin{figure}[t]
\includegraphics[width=8cm]{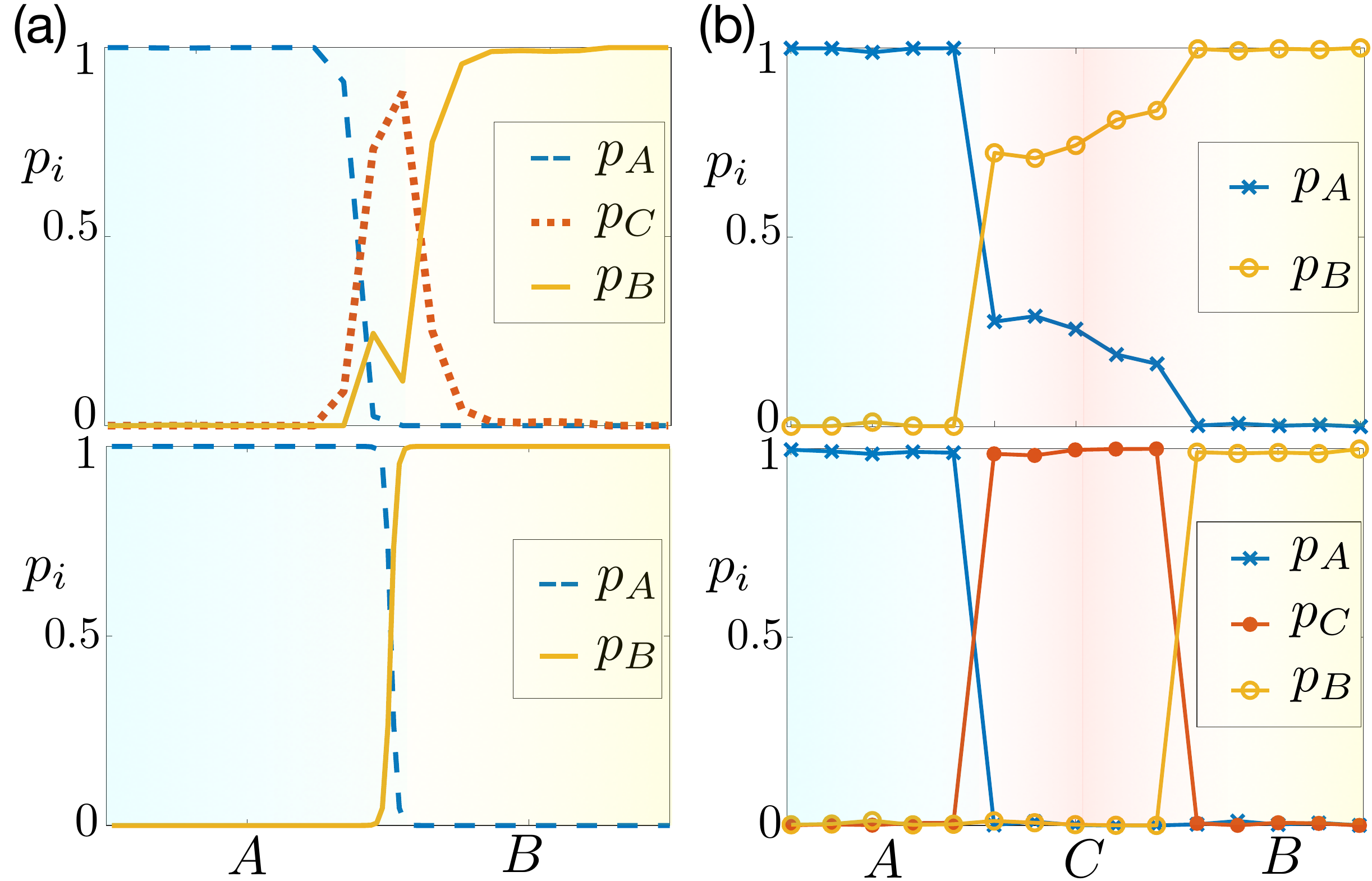}
\caption{Schematic results from applying our NN approach to toy examples identifying (a) a falsely assumed phase C in a system with only two phases A and B, and (b) an unnoticed hidden phase C in a system with three phases A, B, and C.  
(a) shows schematics illustrating situations described in the text, where the upper panel shows the falsely positive result produced by a three-phase classifier for A, B, and C, and the lower panel shows the correct result produced by a two-phase classifier for A and B.
(b) shows results from calculation using MNIST data as input, where we associate A, B, and C with digit $1$, $6$, and $3$, respectively. The upper panel shows the falsely negative result produced by a two-phase classifier for only A and B. The lower panel shows the correct result produced by a three-phase classifier for A, B, and C. Each tick on the horizontal axis corresponds to a group of 160 MNIST images of the associated digit. 
In both plots, $p_i$ is the confidence for identifying certain input data as phase $i=A,B,C$, and the horizontal axis with background color blue, yellow, and red correspond to phase A, B, and C, respectively.\label{mnist}
}
\end{figure}

\textit{Model and method---}
The model we study is the generalized Aubry-Andre (GAA) model~\cite{GAASPME} $H=H_0+H_{int}$ in a 1D system of size $L$, where 
\begin{align}
&H_0=\sum_{j=1}^L \left[-t(c^{\dagger}_jc_{j+1}+H.c.) + 2\lambda\frac{\cos(2\pi q j+\phi)}{1-\alpha\cos(2\pi q j+\phi)}n_j \right], \notag\\
&H_{int}=V\sum_{j=1}^Ln_{j+1}n_j. \label{eq:GAAmodel}
\end{align} 
Here $n_{j}=c^{\dagger}_jc_{j}$ is the fermionic number operator at site $j$, 
$V$ is the nearest-neighbor interaction strength, and 
$t$ is the nearest-neighbor hopping strength as well as the energy unit throughout the article. 
The second term in $H_0$ describes an incommensurate potential with strength $2\lambda$, an irrational wave number $q=2/(1+\sqrt{5})$, a randomly chosen global phase $\phi$, and a dimensionless parameter $\alpha\in(-1,1)$.

The $\alpha=0$ limit of Eq.~\eqref{eq:GAAmodel} corresponds to the pristine AA model~\cite{aubry1980analyticity}, which does not have an SPME. 
For the general $\alpha\neq0$ case, however, an SPME generally exists at $V=0$. Here we choose $\lambda=0.3$ and $\alpha=-0.8$ to achieve a comparable number of localized and extended single-particle orbitals at $V=0$~\cite{XiaoPeng_PRL}. 
As reported in Ref.~\onlinecite{XiaoPeng_PRL}, the interacting many-body spectrum at some fixed $\lambda$ may exhibit the NEM phase in a finite energy window $E_L<E<E_T$, whereas MBL and ETH phases have energies $E<E_L$ and $E>E_T$, respectively. 
The important question we study in this work using our NN approach is the existence of the NEM phase in the interacting GAA model in the intermediate energy window. 

We now describe how we build an $M$-phase classifier based on a candidate phase diagram that contains $M$ phases, which serves as the building block of our NN approach. 
The network structure of the classifier contains an input layer, a hidden layer, and an output layer [Fig.~\ref{GAAPD}(a)]. 
The size of the input layer is determined by the size of the input data, and the hidden layer contains $30$ sigmoid neurons. 
The output layer contains $N$ softmax neurons, and each produces a real number $p_i\in[0,1], i=1,\cdots,M$, with $\sum_{i=1}^Mp_i=1$. 
Thus, each output $p_i$ can be viewed as the confidence the classifier identifies the input data as belonging to the phase $i$. 
We generate the input data by calculating ES of the interacting GAA model using exact diagonalization with a varying global phase $\phi$ and a fixed particle number $N=L/6$.
The training data set for each phase $i$ is generated from one energy bin~\footnote{The generated ES data are sorted into a series of energy bins with a width of $0.04t$, which constitute the unit of energy throughout the rest of this paper. We have checked that the size of these energy bins is sufficiently small for ensuring convergence of our numerics.}
labeled by $E_i$, $(i=1,\cdots,M)$ deep in each phase $i$ according to the assumed phase diagram we have in mind. 

During the training process, we feed the training data to the input layer and allow the all-to-all couplings between the adjacent layers to evolve from randomly chosen initial values according to the loglikelyhood cost function.
We then test the trained network with another independent set of testing data obtained in the same way. 
If the training is successful, which we define as obtaining a testing accuracy over $99\%$, we feed the trained network with ES from all energy bins throughout the spectrum in order to obtain the resulting phase diagram, which contains energy-resolved confidence $p_i(E)$, $i=1,\cdots,M$.

\textit{The NN approach---}
We develop a recursive procedure that consists of systematically building different classifiers starting from a candidate phase diagram to be tested, and telling from the outputs of these classifiers the correct number of phases.
The technique is powerful enough to identify both falsely positive (incorrectly identifying a non-existing phase) and falsely negative (not identifying an existing phase) cases. 
Here we first demonstrate our general approach using a toy example, where we associate each `phase' with a set of images of a hand-written digit from the MNIST database~\cite{mnist}, a canonical source of input datasets for benchmarking machine learning algorithms. 
To better connect to the dynamical phase diagram of interest in this work, we present the results of this example by `phase diagrams' consisting of different digits. To mimic the continuous tuning parameter in usual phase diagrams, we divide the testing data for each digit into groups and plot the output of the network against the group labels.

First imagine a case where the studied phase space contains only phases A and B, but we falsely assume that a phase C exists in between.  
To test our assumption, we start by training a three-phase classifier with data obtained from small regimes within the phase spaces of A, B, as well as a phase space that we thought to be C but is actually a part of B. 
Two scenarios can happen in this case. 
First, for simpler phases with a low variance within each phase, the training procedure itself would fail with low testing accuracy. 
Our example based on MNIST data falls in this category. 
Second, for more complicated phases with a large variance within each phase, the training process could 
be successful but the regime identified as phase C with a high confidence ($p_C\rightarrow1$) would be negligible or (at best) similar in size to the small regime where the training data for C were collected. 
This is because instead of capturing universal properties of a phase, the network is actually trained to capture detailed features tied to the small training regime. 
We show a schematic in the upper panel of Fig.~\ref{mnist}(a) illustrating this second scenario, where the narrowly peaked confidence curve $p_C$ and the apparent confusion between B and C suggest that there are fewer phases in reality than what we assumed. 
This is in contrast to the phase diagram produced by a two-phase classifier for A and B [bottom panel in Fig.~\ref{mnist}(a)] that matches the reality, where each curve exhibits high confidence over a substantial phase space. 

The above guidelines for identifying a falsely assumed phase can be further exploited to identify hidden phases.
Now imagine another case where a phase C exists between phases A and B, but we only know of the latter two. 
To avoid overlooking any hidden phases, we perform the following recursive three-step procedure.
Step I, we train a two-phase classifier for phases A and B. 
Step II, we apply the previous guideline to the resulting phase diagram [top panel of Fig~\ref{mnist}(b)] and find that neither of the confidence curves $p_{A}$ and $p_B$ is narrowly peaked, which indicates that both A and B phases exist.
Step III, we assume some hidden phase C to exist within the regime where neither $p_{A}$ nor $p_{B}$ is high, and build a three-phase classifier accordingly. 
By re-applying step II to the resulting phase diagram [bottom panel of Fig~\ref{mnist}(b)], we again find that all three phases exist. 
When we further repeat step III to build a four-phase classifier assuming some phase D to exist between A and C or between C and B, however, the low testing accuracy in the training process suggests that phase D does not exist. 
We thus conclude that there exist only three phases A, B, and C as shown in the bottom panel of Fig~\ref{mnist}(b). 

\begin{figure}[!]
\includegraphics[width=8cm]{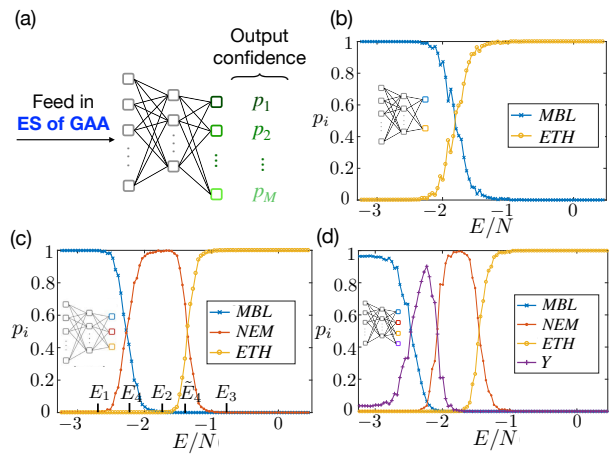}
\caption{(a) The schematics of the building block in our NN approach, a general $M$-phase classifier for phase $i=1\cdots M$. 
The phase diagrams of the GAA model with $L=30$ sites, potential strength $\lambda=0.3$, and interaction strength $V=1$  produced by (b) a two-phase classifier, (c) a three-phase classifier, and (d) a four-phase classifier. Here $p_i(E)$ are the energy-dependent confidence at which the corresponding classifier identifies the eigenstates to be in each of the studied phases. The training data for the classifiers are collected from energy bins $E_1\cdots E_4$ and $\tilde{E}_4$ labeled in (c) for corresponding phases as discussed in the text.
\label{GAAPD}}
\end{figure}

\textit{Results---}
We now employ our NN approach to study the phase diagram of the GAA model in a system with $L=30$ sites, fixed potential strength $\lambda=0.3$, and interaction strength $V=1$. 
First for step I, we assume that the many-body spectrum only consists of MBL near the band edges and ETH in the mid spectrum, and no additional phases in between. 
Based on this assumption, we train a two-phase classifier with data collected from energies deep in MBL ($E_1$) and ETH ($E_3$) phases [Fig. \ref{GAAPD}(c)], respectively.
We find that while the resulting phase diagram [Fig. \ref{GAAPD}(b)] shows two substantial energy regimes identified as MBL and ETH respectively, there is also a substantial regime in between where the network does not show high confidence in identifying it as either. 

Next for step II, we investigate whether there is a third phase X hidden in this transition regime by a three-phase classifier for MBL, ETH, and this phase X which we assume to exist. 
To do so, we train a three-phase classifier for MBL, ETH, and X with data collected from energy bin $E_1$, $E_3$, and $E_2$ [Fig. \ref{GAAPD}(c)], respectively. We then benchmark it against the well-known AA model, 
and find the classifier to be reliable~\cite{SI}.  
Applying this three-phase classifier to the GAA case, we find three substantial energy regimes 
identified as MBL, phase X, and ETH,  respectively with a high confidence as energy increases from the edge to the middle of the spectrum [Fig. \ref{GAAPD}(c)]. 
We emphasize that such a result strongly supports the existence of this third phase X under the lens of ES since (i) the training process is successful with a testing accuracy over $99\%$, and (ii) the width of the identified phase X regime (with over $95\%$ confidence) is seven times wider than the size of the energy bin from which the training data for phase X were produced. c

Before moving on to step III, we first comment on the properties of this phase X. 
First note that the phase diagram obtained from the three-phase classifier using a single diagnostic qualitatively agrees with that obtained using two diagnostics in Ref.~\onlinecite{XiaoPeng_PRL}. 
In particular, the energy range for phase X in our results is slightly smaller but fully contained in that of NEM found in Ref.~\onlinecite{XiaoPeng_PRL}. 
Therefore, the phase X we found here is most likely nonergodic while metallic, hence we will refer to this intermediate phase as NEM in the following. 
Moreover, the ES spectral pattern of the eigenstates we identified as NEM is qualitatively different from that of MBL and ETH states~\cite{SI}. 
This indicates that instead of being merely a mixture of MBL and ETH states over a small energy window, the NEM states are actually a dinstinct type of eigenstates that are distinguishable from the MBL and ETH states by ES patterns. 

Finally for step III, we examine if we overlooked any additional hidden phases in the MBL-to-NEM and NEM-to-ETH transition regimes. 
We first train a four-phase classifier for MBL, NEM, ETH, and a fourth phase Y between MBL and NEM with data collected from energy bins at $E_1$, $E_2$, $E_3$, and $E_4$ respectively [Fig. \ref{GAAPD}(c)]. 
From the resulting phase diagram in Fig. \ref{GAAPD}(d), we find that the confidence curve of phase Y narrowly peaks at the training bin $E_4$, and no energy regime can be identified as phase Y with confidence over $90\%$. These observations suggest that phase Y does not exist, in sharp contrast to the results from the three-phase classifier, where we found a wide energy regime identified as NEM with high confidence [Fig.~\ref{GAAPD} (a)].
We also find that there exists no hidden phases between NEM and ETH by performing a similar calculation replacing phase Y with a phase $\tilde{Y}$ between NEM and ETH, where the training data are collected from $\tilde{E}_4$ \cite{SI}.
Thus we predict that the actual phase diagram is the one produced by the three-phase classifier in Fig.~\ref{GAAPD} (c), which supports the existence of NEM but no additional hidden phases. 

\begin{figure}[!]
\includegraphics[width=8cm]{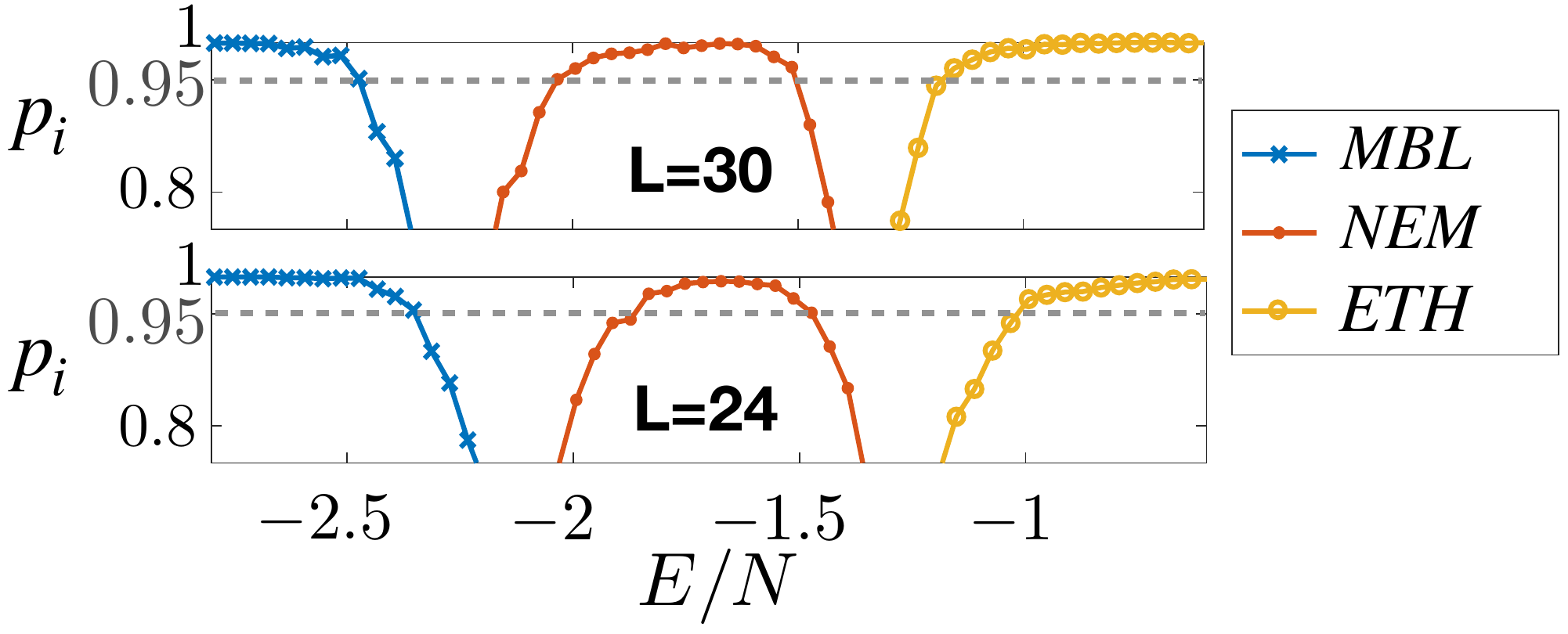}
\caption{
The comparison between the $L=30$ (upper) and $L=24$ (lower) phase diagrams of the GAA model zoomed in on the NEM regimes. Here the parameter choices are the same as those in Fig.~\ref{GAAPD}.  
\label{GAAprop}}
\end{figure}

After establishing the existence of NEM in the $L=30$ system, we further investigate its stability under finite-size effects. 
We build a three-phase classifier for the $L=24$ interacting GAA model, benchmark it against the interacting AA model~\cite{SI}, and use it to study the resulting phase diagram. 
We find that the regime identified as NEM with over $95\%$ confidence increases by $40\%$ as $L$ increases from $24$ to $30$ [Fig. \ref{GAAprop}(b)]. 
Moreover, for the $L=18$ case we even fail to achieve a successful training process under a reasonable hyper-parameter scan.
The above observations are both consistent with the NEM regime becoming more robust as $L$ increases. 
Furthermore, although studying any interacting systems with size $L>30$ is not feasible under our current supercomputer resources, a finite-size analysis in the noninteracting limit can provide additional hints on the robustness of the NEM phase~\cite{SI}. 
Based on our interacting and non-interacting studies, we conclude that it is extremely unlikely that our unbiased identification of the NEM phase as distinct from MBL/ETH phases can be a finite size artifact.

\textit{Conclusion---}
We have developed a neural-network based method for determining the existence of a novel dynamical quantum phase near many-body localization transition, which is an application of supervised machine learning beyond locating phase boundaries among existing phases. Our method allows one to detect hidden phases or conversely identify false hypothetical phases by systematically building different neural-network classifiers. 
By using this technique, we have established that interacting 1D incommensurate systems with single-particle mobility-edges may contain a still-not-very-well-understood nonergodic but metallic phase in the mid energy spectrum. 
Such a phase has an ES spectral pattern very distinct from that of MBL and ETH.    
We have shown that the technique is highly reliable with confidence levels for various identified phases reaching $>99\%$ even using ES data from systems rather modest in size. 
We mention that our method is related to Ref.~\onlinecite{van2017Learning} in that the requirement of \textit{a prior} knowledge of the phase diagram is minimized. 
Meanwhile, our technique focuses on uncovering hidden phases in systems with multiple phases.
Our technique is general, and should be applicable to both equilibrium and nonequilibrium quantum problems.

\emph{Acknowledgment.}---The authors acknowledge helpful discussions with Jordan Venderley, Shenglong Xu, Xiaopeng Li, and Brian Swingle. This work is supported by Microsoft and Laboratory for Physical Sciences. The authors acknowledge the University of Maryland supercomputing resources (http://hpcc.umd.edu) made available for conducting the research reported in this paper. This work is supported by Microsoft and Laboratory for Physical Sciences.
Y.-T.H. and X.L. contributed equally to this work.

%

\maketitle
\begin{center}
{\bf\normalsize{SUPPLEMENTARY MATERIALS}}
\end{center}

\section{I.~~~The phase diagram of AA model studied by the three-phase classifier}
After carrying out our NN approach in the main text, we find that the correct phase diagram comes from the three-phase classifier for MBL, NEM, and ETH phases, trained by data collected from bins labeled by $E_1$, $E_2$, and $E_3$ respectively in Fig.~2~(c) in the main text. 
We now test the reliability of this classifier and benchmark it against the phase diagrams for the well-known $\alpha=0$ case (i.e., AA model). 
The many-body spectra of AA model under strong ($\lambda=4.0$) and weak quasiperiodic potential ($\lambda=0.3$) where no intermediate NEM phase exists are expected to contain only MBL and a majority of ETH eigenstates respectively~\cite{Iyer2013Many-body}. 
For a finite-size system, however, we expect that the edges of the energy spectra are more sensitive to the finite-size effect than the mid-spectra, thus reflecting some spurious nonergodicity. 

\begin{figure}[!]
\includegraphics[width=8cm]{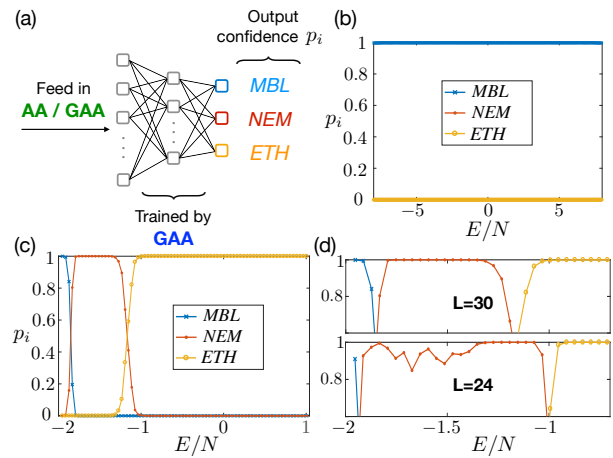}
\caption{(a) The schematics for the three-phase classifier for the MBL, NEM, and ETH phases we benchmark against the AA model. The resulting phase diagrams of the AA model with potential strength (b) $\lambda=4.0$ and (c)-(d)$\lambda=0.3$, where $p_i(E)$, $i=1,2,3$ are the energy-dependent confidence at which the three-phase classifier identifies the eigenstates to be in the MBL, NEM, and ETH phases, respectively. 
The plots in (b) and (c) are for the $L=24$ and $L=30$ case, respectively. 
(d) shows the comparison between the $L=30$ (upper) and $L=24$ (lower) diagrams zoomed in on the NEM regimes. Here we choose the interaction strength $V=1$. 
\label{AAPD}}
\end{figure}

For $\lambda=4.0$, we find that the classifier identifies the eigenstates to be MBL throughout the full energy spectrum, which is consistent with the thermodynamic-limit expectations for the interacting AA model [see Fig. \ref{AAPD}(b)]. 
Moreover, the confidence stays over $99.6\%$ throughout the spectrum already at a smaller system size $L=24$.   
As for the $\lambda=0.3$ case [see Fig. \ref{AAPD}(c)], we find that in a larger system ($L=30$) the classifier identifies eigenstates from most part of the spectrum as thermal and those from a small regime 
near the spectrum edges as non-thermal.
While a non-thermal regime near the spectral edge is consistent with the input data suffering from finite-size effects, we explicitly examine such an explanation by performing the same calculation in a smaller system ($L=24$) and focusing on how the width of the non-thermal regime change as $L$ increases.  
We find that the non-thermal regime shrinks by nearly $40\%$ 
as the system size increases from $L=24$ to $L=30$ [see Fig.~\ref{AAPD}(d)]. 
This finding supports the finite size explanation for the spectral edge states. 
Thus, we conclude that our machine learning method correctly identifies the AA model as not having an intermediate NEM phase. 

\section{II.~~~The phase diagram of GAA model from a four-phase classifier}

\begin{figure}[!]
\includegraphics[width=6cm]{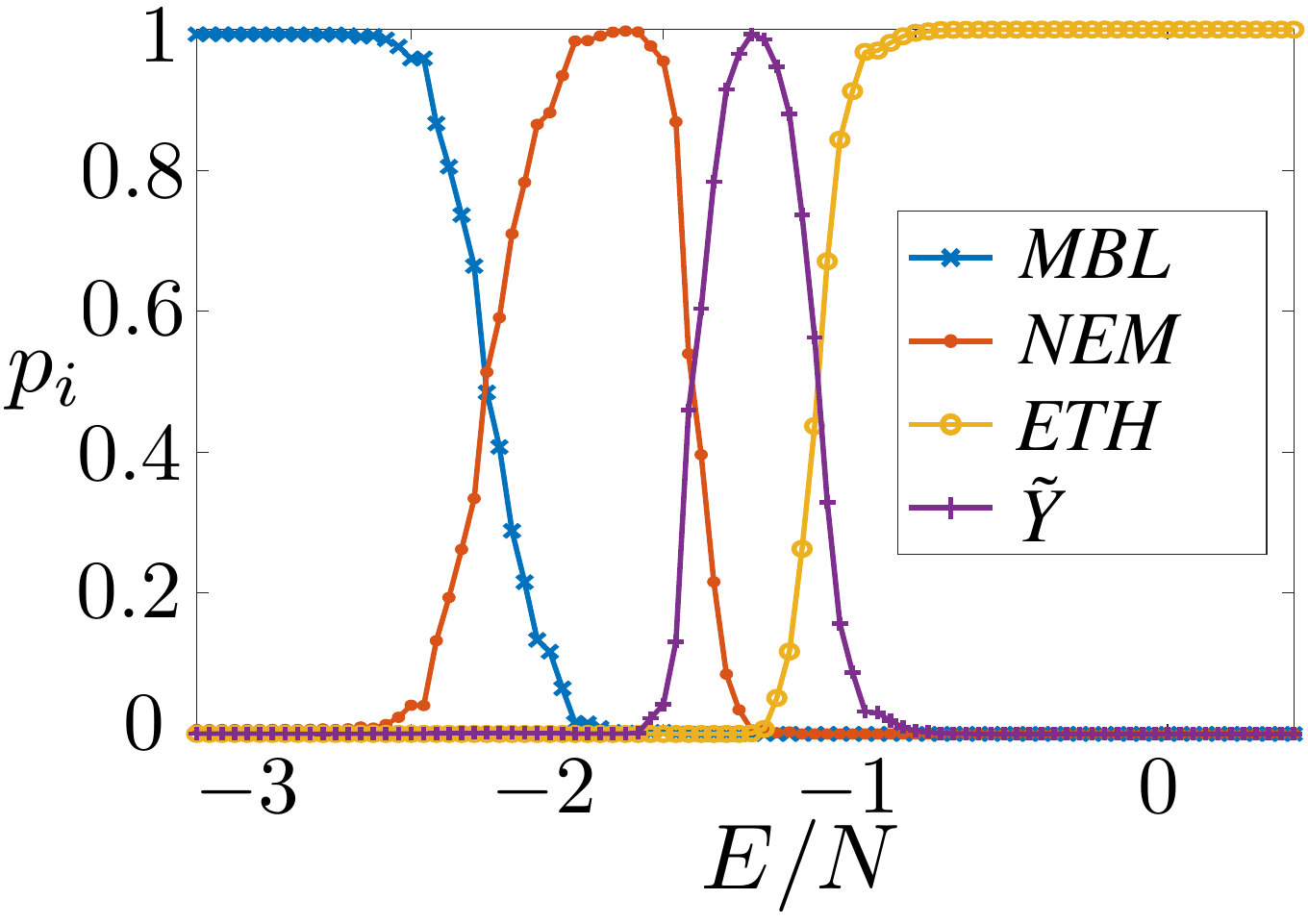}
\caption{The phase diagrams of the GAA model with $L = 30$ sites, potential strength $\lambda = 0.3$, and interaction strength $V = 1$ produced by a four-phase classifier whose training and testing data are collected from the energy bin $\tilde{E}_4$ between the NEM and ETH phases labeled in Fig. 2 (c) in the main text. Here $p_i (E)$ is the energy-dependent confidence at which the corresponding classifier identifies the eigenstates to be in each of the studied phases $i=$MBL, NEM, $\tilde{Y}$, ETH.}
\label{GAA4phase}
\end{figure}
In the main text, we show that there is no hidden phase between MBL and NEM phases using the result of the four-phase classifier for MBL, phase Y, NEM, and ETH [see Fig.~2(d) in the main text]. Here we carry out a similar calculation to examine if there is any hidden phase other than MBL, NEM, and ETH that exists in the NEM-to-ETH transition regime.  
We train a four-phase classifier for MBL, NEM, ETH, and a fourth phase $\tilde{Y}$ between NEM and ETH regimes with data collected from energy bins at $E_1$, $E_2$, $E_3$, and $\tilde{E}_4$ respectively [see Fig. 2(c) in the main text]. 
In the resulting phase diagram [see Fig. \ref{GAA4phase}] we find that although the confidence curve of phase $\tilde{Y}$ does not seem as sharp and low as that of phase $Y$ in the main text, the only bin where it exceeds $99\%$ is exactly the bin $\tilde{E}_4$ from which the training and testing data are collected. 
This indicates that the features of the training data from bin $\tilde{E}_4$ learned by the network are not shared by the neighboring phase space, and thus the phase $\tilde{Y}$ does not exist. Thus, our machine learning protocol correctly avoids false positives. 

\section{III.~~~Finite-size effect from the noninteracting perspective}

For the present problem it is not feasible to study any interacting system with size $L>30$ under currently available supercomputer resources. 
In contrast, as we shown in this section, studying the finite-size effect in the noninteracting limit over a wide range of system sizes $L$ is not only possible but could also provide useful insight for the interacting system with large $L$. 
In particular, we will explain how this finite-size analysis in the noninteracting limit can help us understand the NEM phase found in the interacting GAA model. 

We first briefly explain the main idea. 
For the noninteracting GAA model that possesses an SPME,  
the many-body eigenstates can be classified into three classes according to their `single-partical orbital contents': 
Slater determinants (SDs) consisting of only localized orbitals, only extended orbitals, or both types of orbitals.  
As a result, according to the `orbital contents' of the SDs, the noninteracting many-body spectrum can be divided into three regimes. 
The lowest (highest) energy regime contains only the first (second) type of SDs, whereas the middle part of the spectrum contains all three types of SDs (the `mixed' regime), which are completely mixed in energy. 
Here we conjecture that for the potential strength $\lambda$ and filling fraction $\nu$ we study, the MBL, NEM, and ETH phases found in the weakly interacting GAA model $(V<1)$ are adiabatically connected to these three regimes in the noninteracting limit.
Under such a conjecture, we can speculate on the stability of the NEM phase in a large system from a finite-size analysis of the the mixed regime. (We note as an aside that the inapplicability of this adiabatic conjecture even for a weakly interacting system would imply that the $V=0$ point is singular with the physics of SPME disappearing even for an infinitesimal interaction strength, which, although logically allowed, is not relevant for any physical systems since experimentally the interaction is never strictly zero and the SPME has definitely been experimentally observed\cite{SPME_GAA_Experiment_2018}.

\begin{figure}[!]
\includegraphics[width=8cm]{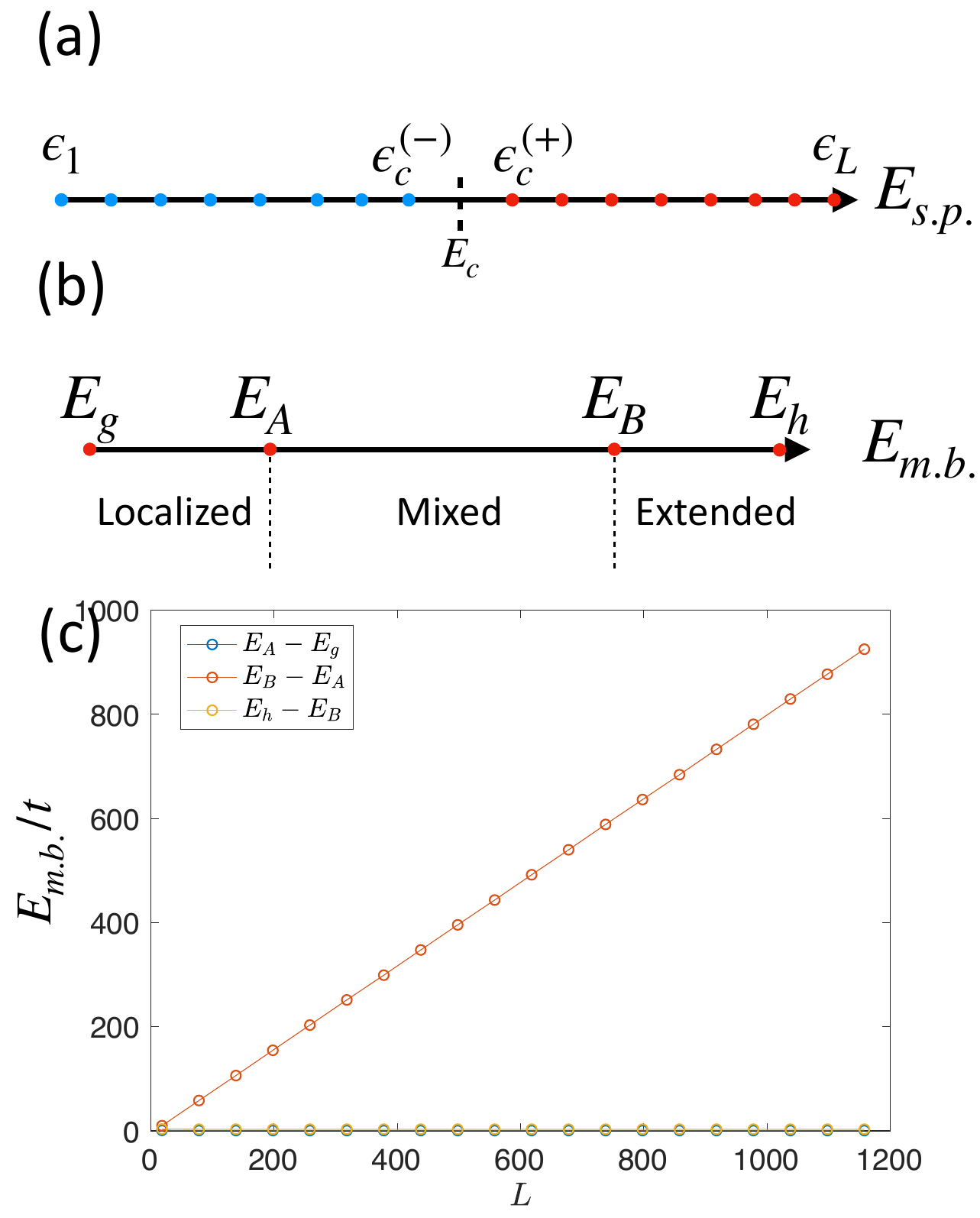}
\caption{
\label{Fig:WidthPlot} 
(a) Illustration of the single-particle (s.p.) energy spectrum. Here blue and red dots represent localized and extended eigenstates, respectively. $\epsilon_c^{(\pm)}$ are the first single-particle eigenstates above and below the mobility edge $E_c$, respectively. 
(b) Illustration of the corresponding many-body (m.b.) energy spectrum made of Slater determinant (SD) states. Two critical energies $E_A$ and $E_B$ mark out the boundaries of the mixed regime (see text). 
(c) Size of the mixed regime in the many-body energy spectrum of the GAA model [given by Eq. (1) in the main text]. The relevant GAA model parameters are $q = 2/(1+\sqrt{5})$, $\phi=0$, $\alpha = -0.8$, $\lambda = 0.3$, and $\nu=1/6$. }
\end{figure}

To explain the above statements in details, we start with a general noninteracting 1D system of $L$ lattice sites (with $L$ single-particle orbitals in total) and a single-particle mobility edge at $E_c$, where there are $M$ localized orbitals below $E_c$ and $L-M$ extended orbitals above $E_c$. 
We further assume that the fraction of localized orbitals $\eta\equiv M/L$ is a constant as $L\to \infty$. 
For concreteness, we label the energies of the single-particle orbitals as $\epsilon_1 < \epsilon_2 < \cdots < E_c < \cdots < \epsilon_L$ (note that $E_c$ is often not a part of the energy spectrum), as shown in Fig.~\ref{Fig:WidthPlot}(a).  

We then construct SDs out of these single-particle orbitals, which are the eigenstates of the noninteracting many-body Hamiltonian. 
Specifically, we assume a total of $N$ states in the system, and keep the filling factor $\nu\equiv N/L$ constant as $L\to\infty$. 
As mentioned earlier, because of the existence of the SPME in the noninteracting GAA model, there are three types of SD states with different `orbital contents', ones consisting of localized orbitals only, ones consisting of extended orbitals only, and ones consisting of both extended and localized orbitals. 
Here we consider the case where the many-body spectrum consists of three energy regimes that differ from each other by the SD types therein,  as this is the case discussed in the main text. The lowest (highest) energy regime contains only the first (second) type of SDs, which we label as the localized (extended) regime, whereas the middle energy regime contains all three types of SDs, which we label as the mixed regime [see Fig.~\ref{Fig:WidthPlot}(b)].
In order for the many-body spectrum to contain all three regimes, the filling factor $\nu$ must satisfy the following condition, 
\begin{align}
	\nu < \min\left\{\dfrac{1}{2}, \eta, 1-\eta\right\}. 
\end{align}
In the example we studied in the main text, the system has $\eta\simeq 0.28$, and the filling factor we choose is $\nu=1/6$. 

We now show that {the mixed regime comprises} the majority of the full many-body energy spectrum in the $L\to\infty$ limit. 
To begin with, we note that the spectrum of SD states is bounded by $E_g$ and $E_h$, 
\begin{align} 
	E_g = \sum_{k=1}^{N} \epsilon_k, \quad E_h = \sum_{k=L-(N-1)}^{L} \epsilon_k, 
\end{align}
which are the lower and higher edges in the many-body energy spectrum, respectively. 
In addition, we can introduce two critical many-body energies $E_A$ and $E_B$ that define the three energy regimes~\cite{XiaoPeng_PRB}, with 
\begin{align}
	E_A = \epsilon_c^{(+)} + \sum_{k=1}^{N-1} \epsilon_k, \quad 
	E_B = \epsilon_c^{(-)} + \sum_{k=L-(N-2)}^{L} \epsilon_k. 
\end{align}
In the above equation, $\epsilon_c^{(+)}$ and $\epsilon_c^{(-)}$ are the energies of the first single-particle orbital above and below $E_c$, respectively. 
In other words, the localized regime lies within $E\in [E_g,E_A)$, the mixed regime lies within $E\in (E_A,E_B)$, and the extended regime lies within $E\in (E_B,E_h]$. 
As a result, the width of the three regimes are given by
\begin{align}
	E_A - E_g &= \epsilon_c^{(+)} - \epsilon_N, \notag\\
	E_h - E_B &= \epsilon_{L-N+1} - \epsilon_c^{(-)}, \notag\\
	E_B - E_A &= \sum_{k=L-N+2}^{L}\epsilon_k - \sum_{k=1}^{N}\epsilon_k +  \epsilon_c^{(+)} - \epsilon_c^{(-)}. 
\end{align}
Clearly, both $E_A - E_g$ and $E_h - E_B$ are intensive quantities while $E_B - E_A$ is an extensive quantity. 
Therefore, the mixed regime will eventually dominate the many-body spectrum as the system size increases. 
Figure~\ref{Fig:WidthPlot}(c) shows a numerical result of the width of the mixed regime in the noninteracting GAA model, and we indeed find that the mixed regime becomes the dominant part as $L$ increases. 

Our conjecture and analysis provide a possible physical picture and hints about the fate in large systems of the MBL, NEM, and ETH phases we found in the interacting GAA model. 
If this conjecture turns out to be true (at least for weak to modest interactions), then based on our analysis in this section, the NEM phase is unlikely to be a finite size artifact. 
However, its eventual existence (or not) in the thermodynamic limit will need to be established in future studies. 
This last statement is in fact equally true for essentially all numerically-found MBL phases too---we do not know their fate  with absolute certainty in the thermodynamic limit.  All we can say is that we have established the existence of the NEM phase at the same level of numerical certainty as the MBL phase in the intreacting GAA model.
\\
\section{IV.~~~Entanglement spectra}
\begin{figure}[t]
\includegraphics[width=8cm]{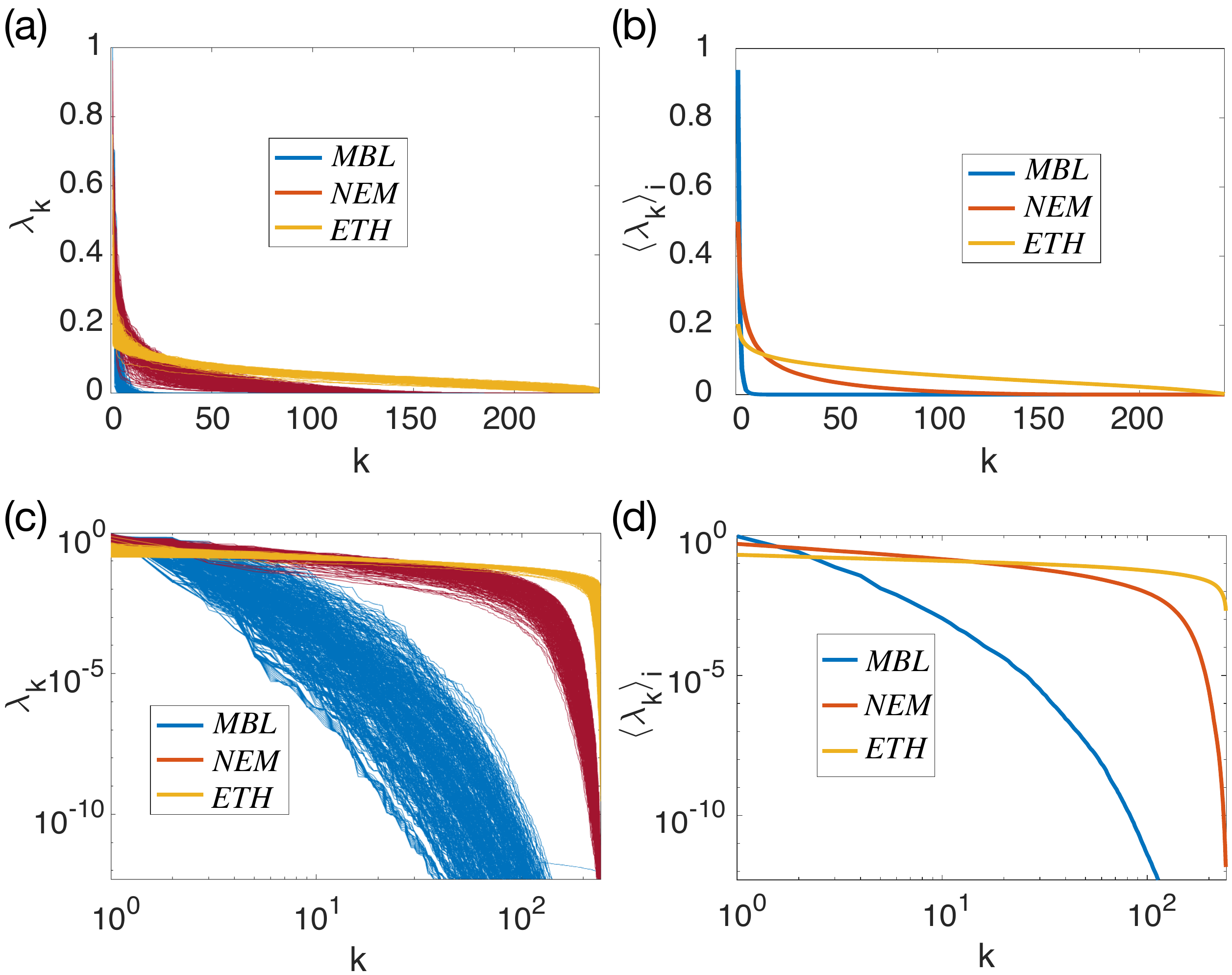}
\includegraphics[width=8cm]{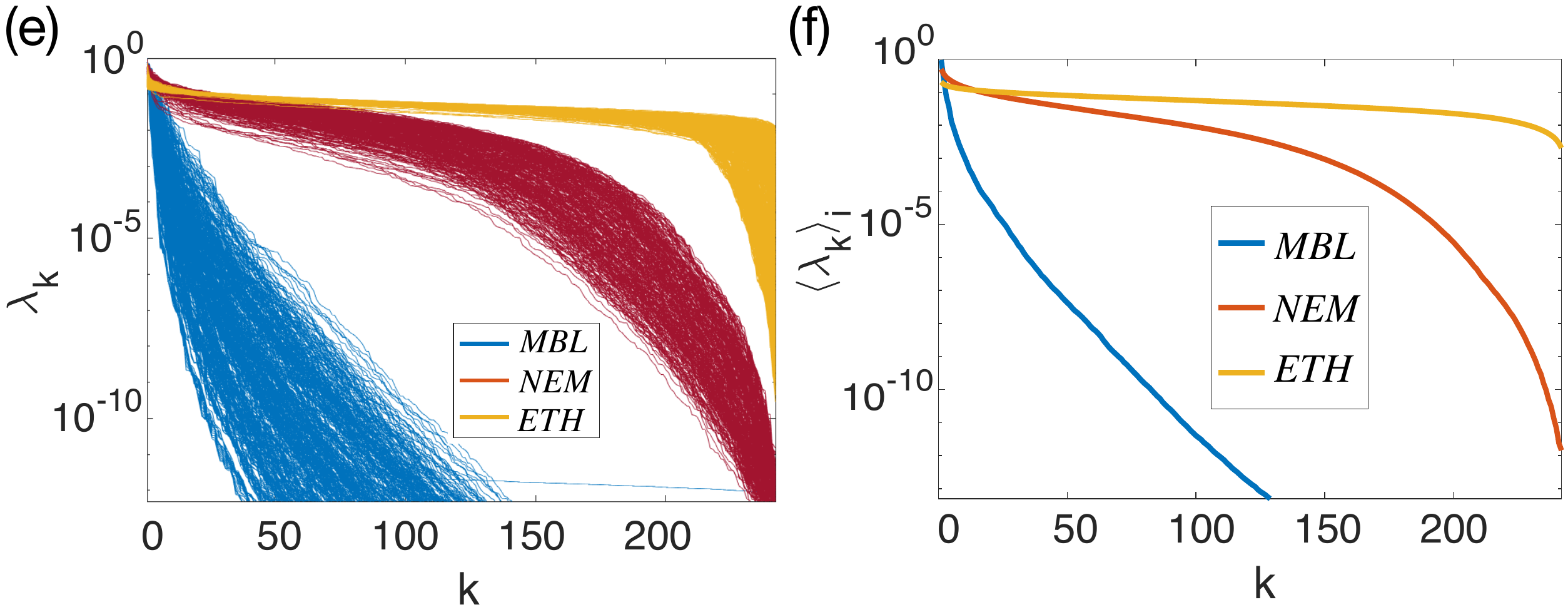}
\caption{(a)(c)(e) The entanglement spectra $\lambda_k$ defined in the text and (b)(d)(e) their averages $\langle\lambda_k\rangle_i$ of randomly chosen eigenstates from all energy bins identified as one of the three phases with $>99\%$ confidence. Here $k$ is the label of the eigenvalues as we arrange them in the decreasing order. We choose 50 random eigenstates from each energy bin, and the average $\langle\cdots\rangle_i$ is over all 50 eigenstates and all energy bins that are identified as phase $i=$ MBL, NEM, ETH. (c)-(d) and (e)-(f) show the log-log and log-linear plots of (a)-(b) respectively.}
\label{ES}
\end{figure}

We present some details about our calculation of the entanglement spectrum data. 
In general for a pure many-body state $\ket{\psi}$, we can carry out a Schmidt decomposition as follows~\cite{Haldane2008PRL}, 
\begin{align}
	\ket{\psi} = \sum_{i} e^{-\xi_i/2}\ket{\chi_A^{i}}\otimes \ket{\chi_B^{i}}, 
\end{align}
where $\{\ket{\chi_A^{i}}\}$ and $\{\ket{\chi_B^{i}}\}$ are the complete set of basis states in the subsystem $A$ and $B$, respectively. 
Formally, the $\lambda_i$'s constitute the entanglement spectrum of the many-body eigenstate $\ket{\psi}$. 
For convenience, we 
re-arrange $\{\lambda_i\equiv e^{-\xi_i/2}\}$ (the singular values of the matrix form of the many-body eigenstate $\ket{\psi}$) in a decreasing order and denote them $\{\lambda_k\}$. We then use this entanglement spectral pattern $\{\lambda_k\}$ as the input data for the classifier. 

Inspired by how well the network can distinguish the three phases via the entanglement spectral pattern $\lambda_k$, here we directly examine the $\lambda_k$ for all three phases. 
Specifically, for each energy bin with a high confidence level ($p_i>99\%$) in Fig.~2 (c) in the main text, we randomly choose $50$ eigenstates and plot their ES in Fig.~\ref{ES}(a). 
In order to study the qualitative differences between the three phases, we study the averages $\langle\lambda_k\rangle_i$ of all the ES data from the same phase $i$ 
and plot the results in Fig.~\ref{ES}(b)].  
Note that all eigenstates within the same energy bin are produced by varying the global phase $\phi\in[0,2\pi)$ in the GAA model, and they all have slightly different energies. 

From the results in Fig.~\ref{ES}, we make the following observations. 
First, we find that for all eigenstates within the same phase, their $\lambda_k$ share a similar dependence in the index $k$. 
In contrast, 
the $\lambda_k$'s of the eigenstates from  different phases have qualitatively different $k$ dependencies. 
We emphasize that $\lambda_k$ is a property of a single eigenstate, rather than an average over energy or over the global phase $\phi$. 
Therefore, this observation suggests that instead of being merely an average of MBL and ETH eigenstates over a finite energy window, eigenstates from the NEM energy bins are indeed a third type of eigenstates that has a distinct ES pattern than those of MBL and ETH. 
Second, when we consider the averaged ES data, we find that the $k$ dependences of $\langle\lambda_k\rangle_\text{MBL}$ and $\langle\lambda_k\rangle_\text{ETH}$ [see Fig. \ref{ES}(c)-(f)] agree with existing results in the literature~\cite{powerlawES,YangPRL2015}. 
By contrast, we find that $\langle\lambda_k\rangle_\text{NEM}$ clearly has a distinct $k$ dependence, different from that of both MBL and ETH phases. 
Obtaining the explicit form of the $\langle\lambda_k\rangle_\text{NEM}$ is out of the scope of this work and should be the subject of future studies. 

\end{document}